\pdfoutput=1

 \documentclass[twocolumn,trackchanges]{aastex631} 

 \hypersetup{linkcolor=blue,citecolor=blue,filecolor=blue,urlcolor=blue}

\accepted{31 October, 2023}

\usepackage{graphicx}
\usepackage{amsmath}  
\usepackage{xcolor}    
\usepackage{subfigure}

\shortauthors{Sextl et al.}
\shorttitle{TYPHOON: The barred spiral NGC 1365}
\graphicspath{{./}{figures/}}
\begin{document}

\title{The TYPHOON stellar population synthesis survey: \\ I. The young stellar population of the Great Barred Spiral NGC 1365}

\correspondingauthor{Eva Sextl}
\email{sextl@usm.lmu.de}

\author{Eva Sextl}
\affiliation{Universit\"ats-Sternwarte, Fakult\"at f\"ur Physik, Ludwig-Maximilians Universit\"at M\"unchen, Scheinerstr. 1, 81679 M\"unchen, Germany}

\author{Rolf-Peter Kudritzki}
\affiliation{Universit\"ats-Sternwarte, Fakult\"at f\"ur Physik, Ludwig-Maximilians Universit\"at M\"unchen, Scheinerstr. 1, 81679 M\"unchen, Germany}
\affiliation{Institute for Astronomy, University of Hawaii at Manoa, 2680 Woodlawn Drive, Honolulu, HI 96822, USA}
\author{Andreas Burkert}
\affiliation{Universit\"ats-Sternwarte, Fakult\"at f\"ur Physik, Ludwig-Maximilians Universit\"at M\"unchen, Scheinerstr. 1, 81679 M\"unchen, Germany}
\author{I-Ting Ho}
\affiliation{Max-Planck-Institute for Astronomy, K\"onigstuhl 17, D-69117 Heidelberg, Germany}
\author{H. Jabran Zahid}
\affiliation{Microsoft Research, 14820 NE 36th St, Redmond, WA 98052, USA}
\author{Mark Seibert}
\affiliation{The Observatories, Carnegie Institution for Science, 813 Santa Barbara Street, Pasadena, CA 91106, USA}
\author{Andrew J. Battisti}
\affiliation{Research School of Astronomy and Astrophysics, Australian National University, Canberra, ACT 2611, Australia}
\affiliation{ARC Centre of Excellence for All Sky Astrophysics in 3 Dimensions (ASTRO 3D), Australia}
\author{Barry F. Madore}
\affiliation{The Observatories, Carnegie Institution for Science, 813 Santa Barbara Street, Pasadena, CA 91106, USA}
\author{Jeffrey A. Rich}
\affiliation{The Observatories, Carnegie Institution for Science, 813 Santa Barbara Street, Pasadena, CA 91106, USA}


\begin{abstract}

We analyze TYPHOON long slit absorption line spectra of the starburst barred spiral galaxy NGC 1365 obtained with the Progressive Integral Step Method covering an area of 15 square kpc. Applying a population synthesis technique, we determine the spatial distribution of ages and metallicity of the young and old stellar population together with star formation rates, reddening, extinction and the ratio R$_V$ of extinction to reddening. We detect a clear indication of inside-out growth of the stellar disk beyond 3 kpc characterized by an outward increasing luminosity fraction of the young stellar population, a decreasing average age and a history of mass growth, which was finished 2 Gyrs later in the outermost disk. The metallicity of the young stellar population is clearly super solar but decreases towards larger galactocentric radii with a gradient of -0.02 dex/kpc. On the other hand, the metal content of the old population does not show a gradient and stays constant at a level roughly 0.4 dex lower than that of the young population. In the center of NGC 1365 we find a confined region where the metallicity of the young population drops dramatically and becomes lower than that of the old population. We attribute this to infall of metal poor gas and, additionally, to interrupted chemical evolution where star formation is stopped by AGN and supernova feedback and then after several Gyrs resumes with gas ejected by stellar winds from earlier generations of stars. We provide a simple model calculation as support for the latter.

\end{abstract}



\section{Introduction} \label{sec:intro}

Spectroscopic studies of galaxies with integral field units (IFU) have become an important tool to investigate the evolution of galaxies. Spatially resolved maps of their chemical composition, stellar ages, star formation rates and gas properties provide unique information about the complex physical processes affecting galaxy evolution (see, for instance, \citet{bittner2020}, \citet{Carrillo2020}, \citet{parikh2021}, \citet{Emsellem2022}, \citet{Pessa2023}, \citet{westmoquette2011}, \citet{sanchez2014},\citet{Thaina2023}). Usually, because of the relatively small field of view of the available IFUs (of the order of one arcminute) this work has been mostly concentrated on galaxies with small angular size or only central regions. Therefore, given the enormous potential of these spectroscopic stellar population studies, we have started a project with the goal to extend this work to cover large parts of galactic disks together with their central regions. The TYPHOON survey which uses stepwise combined long slit spectra of galaxies seems ideal for this purpose. For the population synthesis analysis of the integrated stellar population absorption line spectra we will use the technique developed and described in \citet{Sextl2023}.

We start the project with the population synthesis analysis with the starbursting type 2 Seyfert barred-spiral NGC 1365 in the Fornax cluster. With an isophotal radius of R$_{25}$ = 5.61 arcmin or 29.55 kpc at a distance of 18.1 Mpc \citep{Ho2017} NGC 1365 is a galaxy of huge dimensions and ideally suited for a long-slit IFU-like investigation. It is well studied with respect to the dynamics of its gas and stellar content (\citet{Lindblad1999}, \citet{Sanchez2008}, \citet{Jalocha2010}). It is an almost face-on galaxy, which avoids line of sight confusion complications and minimizes effects of interstellar extinction. The central region shows extensive star formation and the presence of a low luminosity AGN with two conical outflows (see \citet{venturi2018} and references therein).
In addition, the chemical composition of its ionized ISM has been carefully investigated \citep{Ho2017, Chen2023} and first long slit star formation studies have already been carried out \citep{sanchezBlazquez2011}. Since the galaxy is characterized by strong star formation activity, the blue sensitivity of TYPHOON is an advantage and allows for a good characterization of the properties of its young stellar population.  This will provide important information extending the work and the results obtained recently within the comprehensive ESO VLT PHANGS-MUSE survey \citep{Emsellem2022,Pessa2023}. 

We describe the observations in section \ref{sec:obs} and the population synthesis analysis technique in section \ref{sec:analysis}. The results are presented in sections \ref{sec:results} followed by a discussion in \ref{sec:discussion}.

\begin{figure}[htb!]
	\center \includegraphics[width=1\columnwidth]{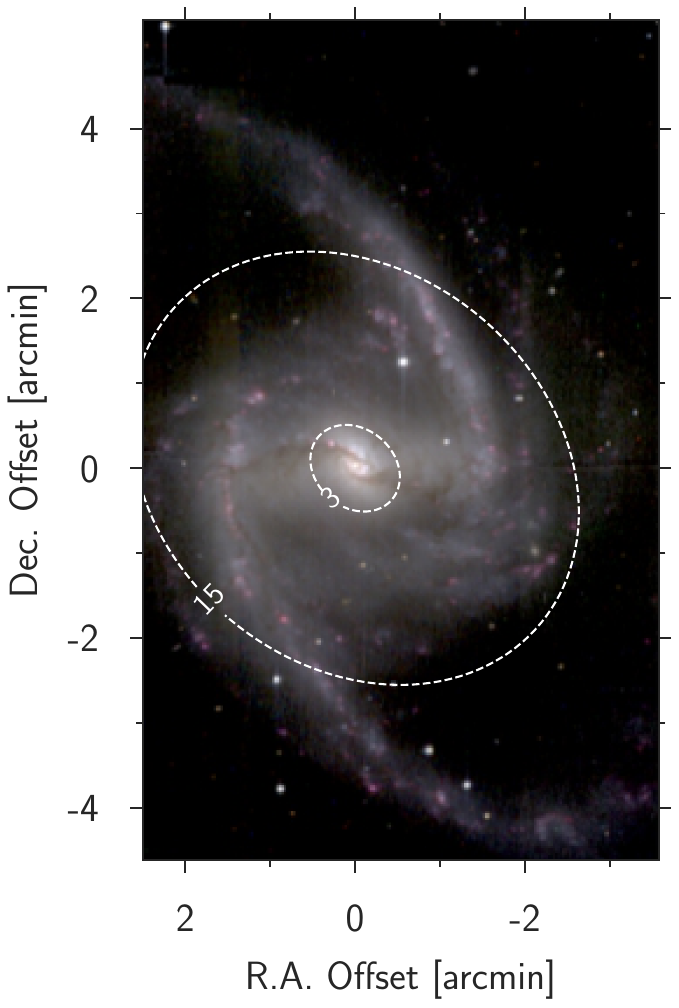}\medskip
	\caption{TYPHOON BVI color composite image of NGC 1365. The position of the central AGN is indicated by a small green cross. The bar extends roughly 200" on the sky \citep{Lindblad1999} corresponding to $\sim$17.6 kpc projected length. The two dashed ellipses indicate galactocentric distances of 3 and 15 kpc, respectively. East is towards the left and north towards the top.}    \label{fig:BVR}
\end{figure}

\section{Observations} \label{sec:obs}

The TYPHOON survey (P.I. B. Madore) uses the Las Campanas du Pont 2.5m telescope Wide Field CCD imaging spectrograph with a custom long-slit of 18 arcmin length and 1.65 arcsec width which progressively scans across the galaxies (Progressive Integral Step Method, PrISM) to construct 3D data cubes of 1.65 times 1.65 arcsec$^2$ spaxels. At a distance of 18.1 Mpc \citep{Jang2018} 1.65 arcsec are equivalent to 145 pc. The spectral resolution corresponds to a FWHM of 8.2 \AA. In the case of NGC 1365 the slit was placed along the north-south direction. More details of the observations are described in \citet{Ho2017} and \cite{Chen2023}. Figure \ref{fig:BVR} provides a BVR color composite image constructed from the TYPHOON data cube. The figure has already been shown in  \citet{Ho2017} but is repeated here for illustration. NGC 1365 is a massive galaxy with a stellar mass of log M$_*$ = 10.95 (in solar units, \citealt{Munoz2013}, \citealt{Leroy2019}), an isophotal radius of 5.61 arcmin \citep{deVaucouleurs1991}, inclination angle of 35.7 degrees and a position angle of 49.5 degrees based on 2MASS photometry (\citealt{Jarrett2003}, see \citealt{Ho2017} Table 1). 

For our population synthesis analysis we use the spectral range from 4000 to 7070 \AA~ where the flux calibrated spectra have the best signal. Compared to the range from 4800 to 7000 \AA~ used in the PHANGS-MUSE study by \citet{Pessa2023} this is an important blueward extension which enables a more accurate characterization of the young stellar population.

\begin{figure}[htb!]
       \medskip
	\center \includegraphics[width=1\columnwidth]{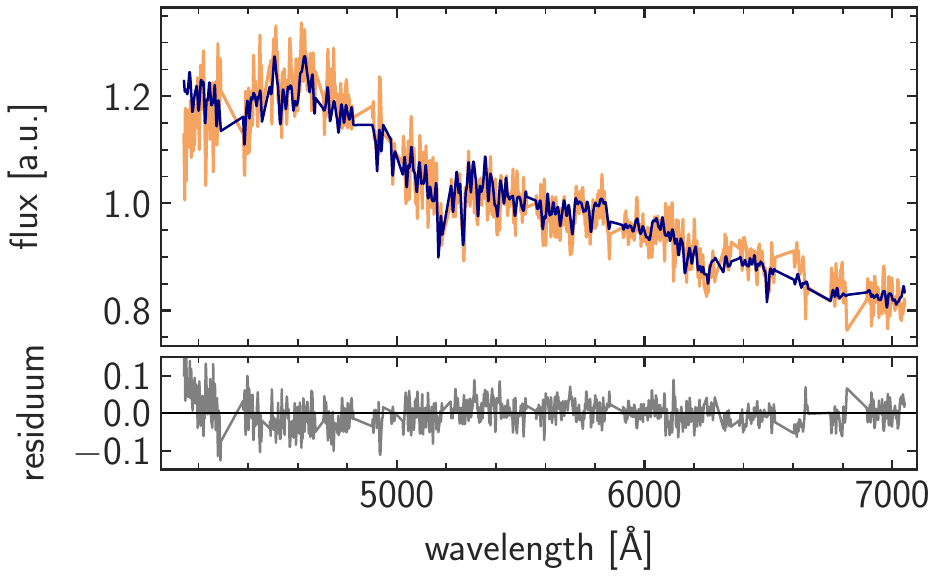}\medskip
    \center \includegraphics[width=1\columnwidth]{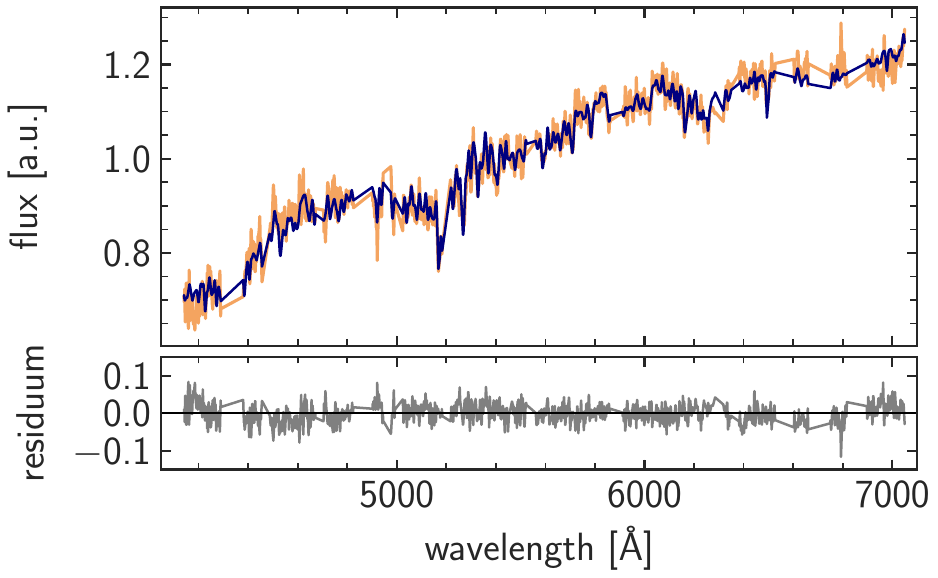}\medskip
	\caption{TYPHOON spectrum (orange) and the corresponding stellar population fit (dark blue). Note that for the fit ISM emission and absorption lines and broad stellar Balmer lines are masked out. The resulting gaps are shown as straight lines. The top spectrum corresponds to bin 16 at $\Delta\alpha = -1.073$ arcmin and $\Delta\delta = 1.292$ arcmin (northern spiral arm) and the bottom spectrum to bin 153 at $\Delta\alpha = -0.192$ arcmin and $\Delta\delta = -0.11$ arcmin (on the southern dust lane in the center). The different physical properties of the stellar populations are discussed in section \ref{sec:results}.}     \label{fig:specfitT}
\end{figure}

\begin{figure}[htb!]
	\center \includegraphics[width=0.96\columnwidth]{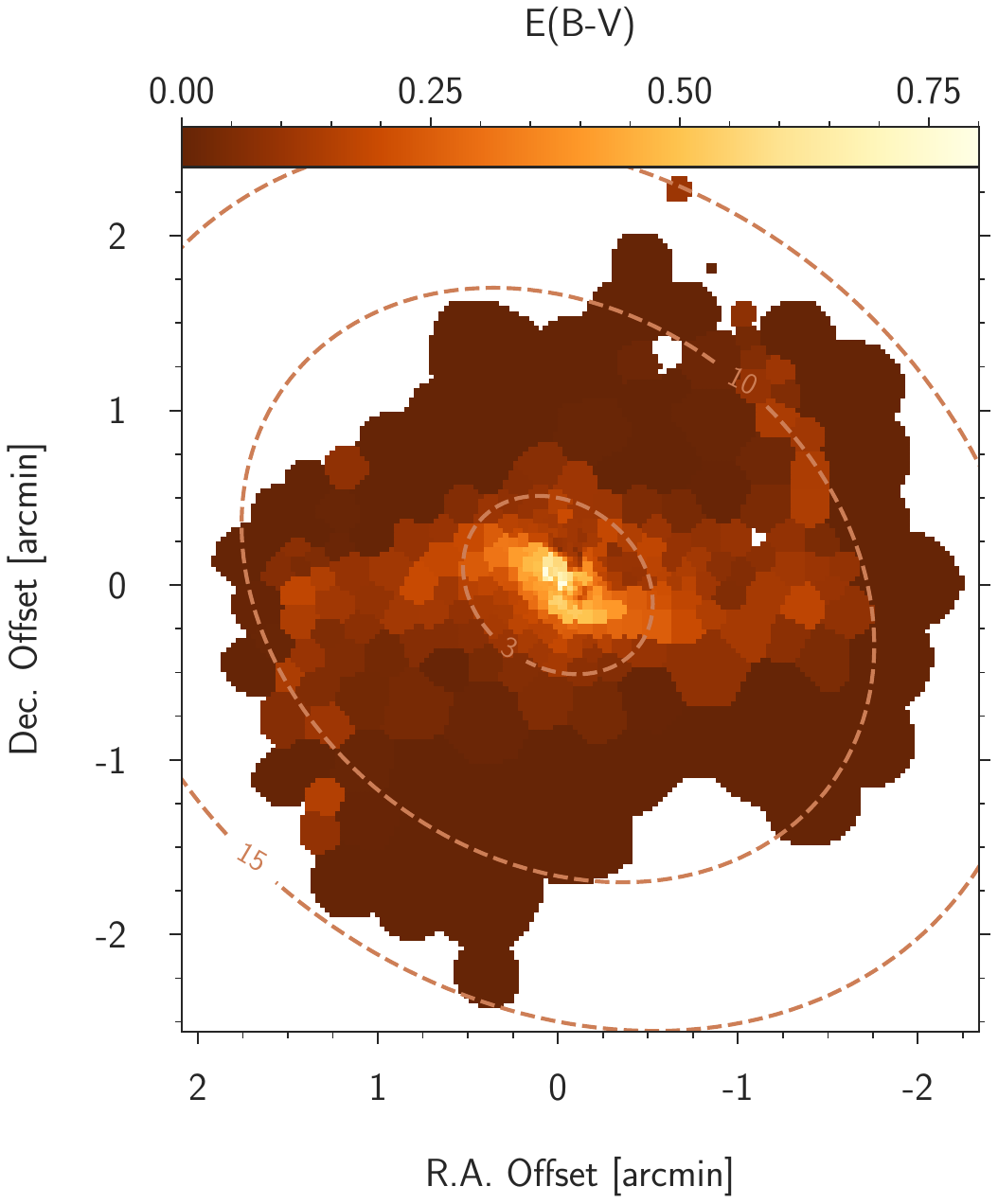}\medskip
    \center \includegraphics[width=0.96\columnwidth]{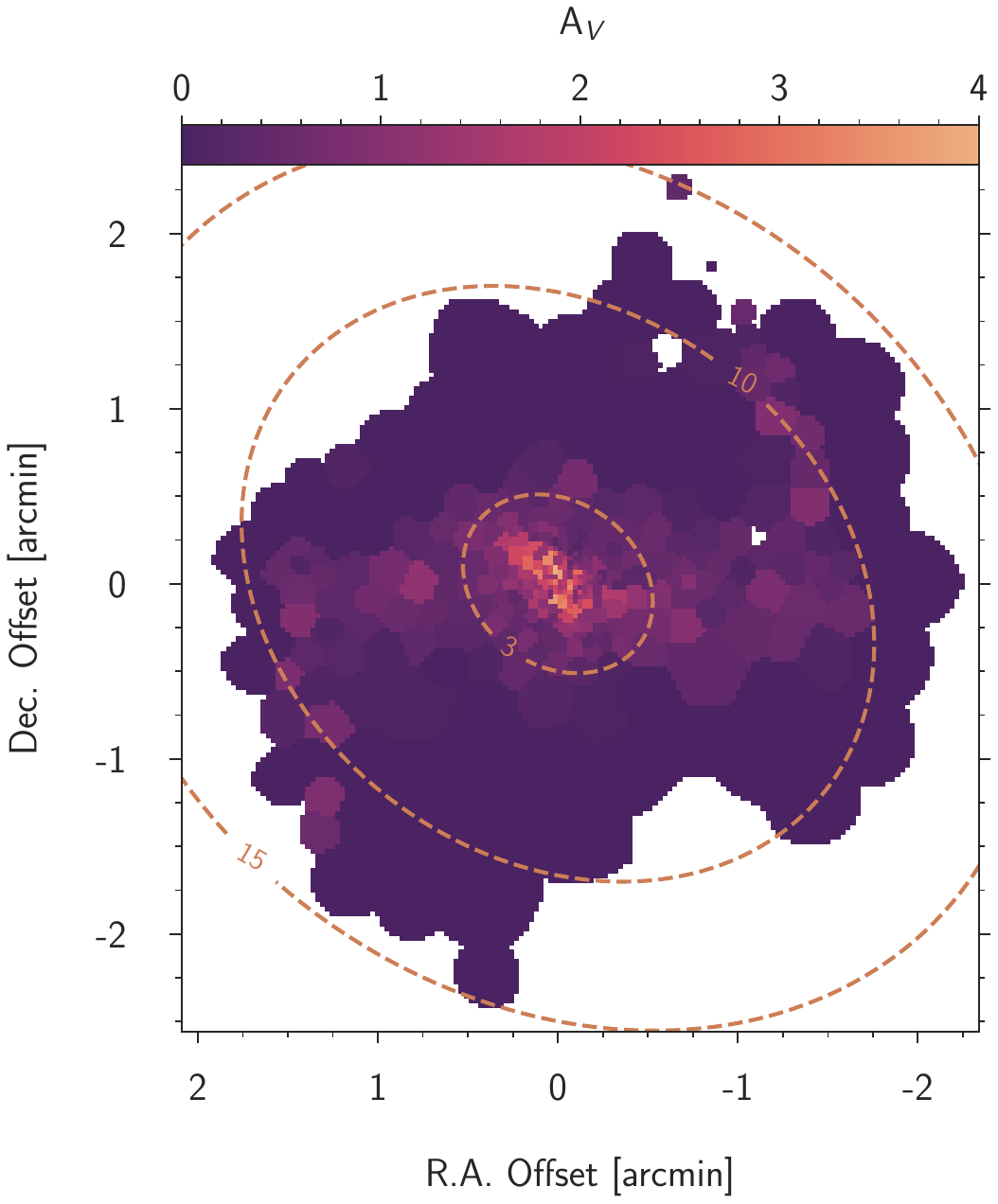}\medskip
	\caption{Reddening (top) and extinction (bottom) maps of NGC 1365 obtained with our population synthesis fit. The central region is indicated by the green cross. The ellipses indicate galactocentric distances of 3, 10, and 15 kpc.}     \label{fig:redden}
\end{figure}

If needed, we combine the spectra of neighboring individual spaxels by Voronoi binning using \citet{cappellari2003} to obtain a minimum signal-to-noise ratio of 30 in the stellar continuum at 5000 \AA. However, in order to avoid averaging over too large spatial dimensions we exclude bins consisting of more than 400 TYPHOON spaxel. We also remove bins containing the contribution of bright foreground stars. This leaves us with spectra of 359 bins distributed over the galaxy. As an example, Figure \ref{fig:specfitT} shows the spectra of two bins and the corresponding stellar population fits. Figures \ref{fig:redden} and \ref{fig:agemap} give an impression of the spatial distribution of the bins analyzed.

\begin{figure}[htb!]
\medskip
	\center \includegraphics[width=1\columnwidth]{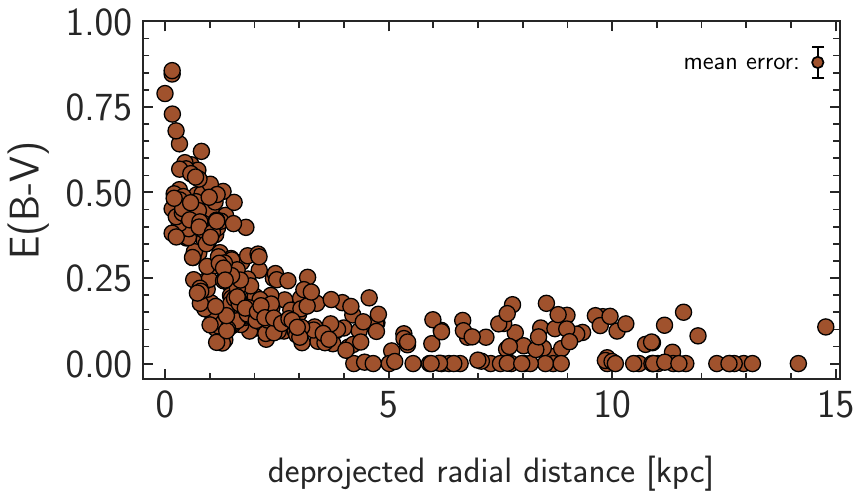}\medskip
    \center \includegraphics[width=1\columnwidth]{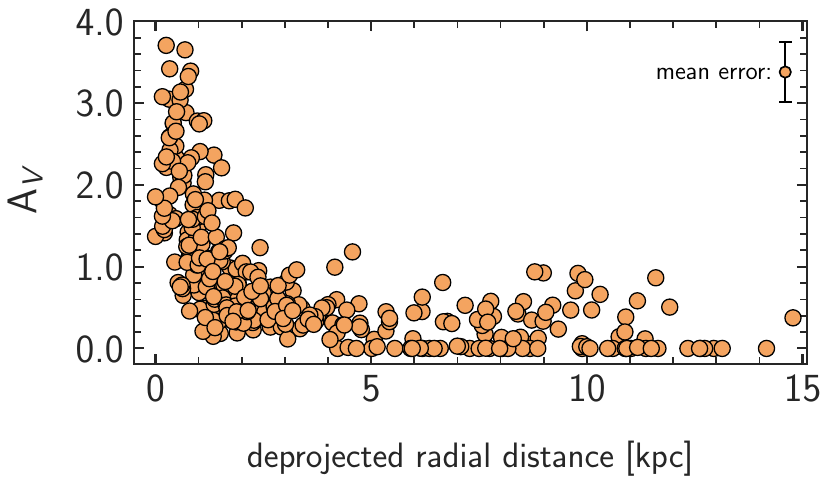}\medskip
    \center \includegraphics[width=1\columnwidth]{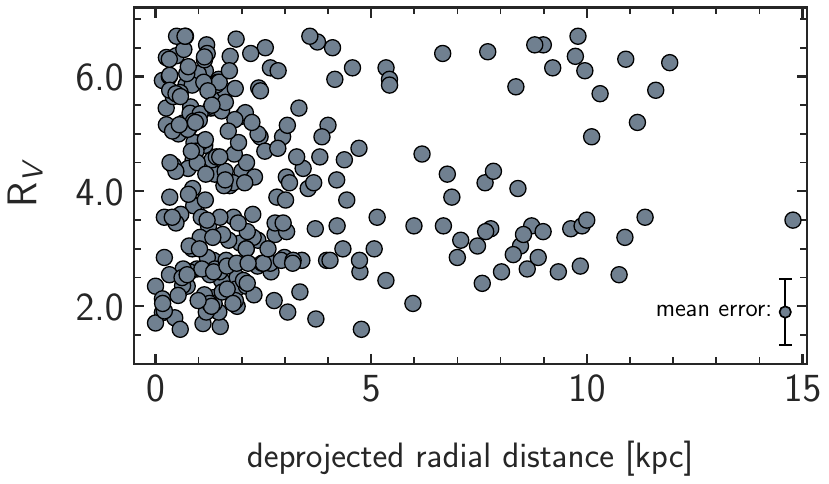}\medskip
	\caption{Radial distribution as a function of deprojected galactocentric distance of reddening E(B-V) (top), extinction A$_V$ (middle) and R$_V$ (bottom).}     \label{fig:redgrad}
\end{figure}


\begin{figure}[htb!]
	\center \includegraphics[width=1\columnwidth]{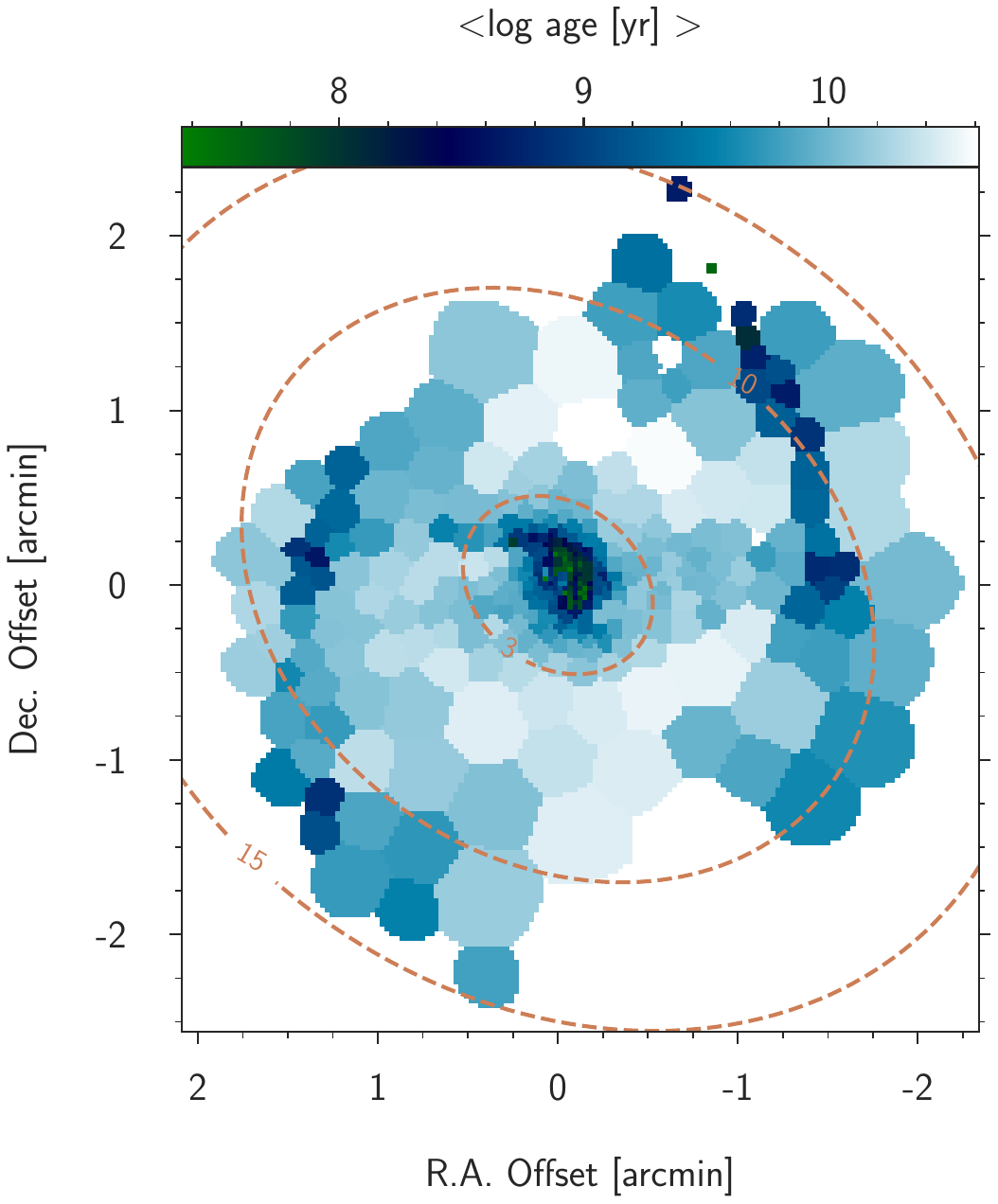}\medskip
	\caption{Map of average population age. The dashed ellipses indicate galactocentric distances of 3, 10, and 15 kpc, respectively.}     \label{fig:agemap}
\end{figure}

\begin{figure}[htb!]
    \center \includegraphics[width=1\columnwidth]{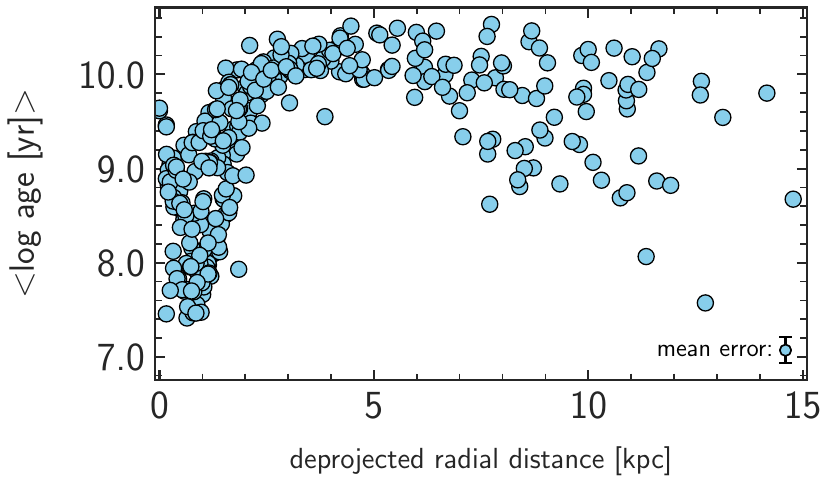}
    \center \includegraphics[width=1\columnwidth]{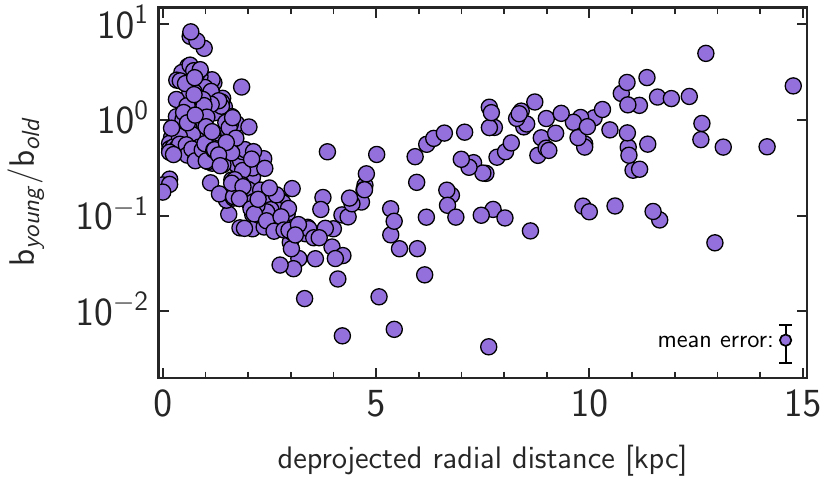}\medskip
    \center \includegraphics[width=1\columnwidth]{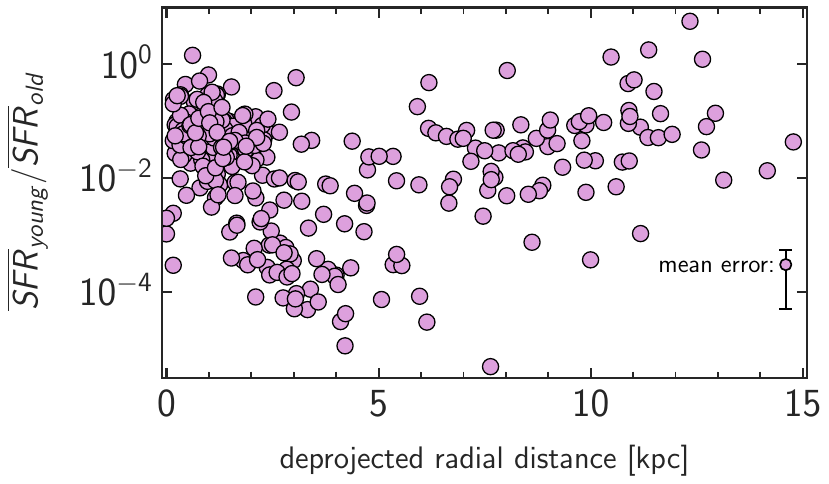}\medskip
	\caption{Radial distribution of average population age (top), the ratio b$_{young}$/b$_{old}$ (middle), which is ratio of the contribution of the young and old population to the total observed population spectrum, and the ratio of average star formation rates of the young to the old population.}     \label{fig:agegrad}
\end{figure}

\begin{figure*}[htb!]
	\center \includegraphics[width=1.5\columnwidth]{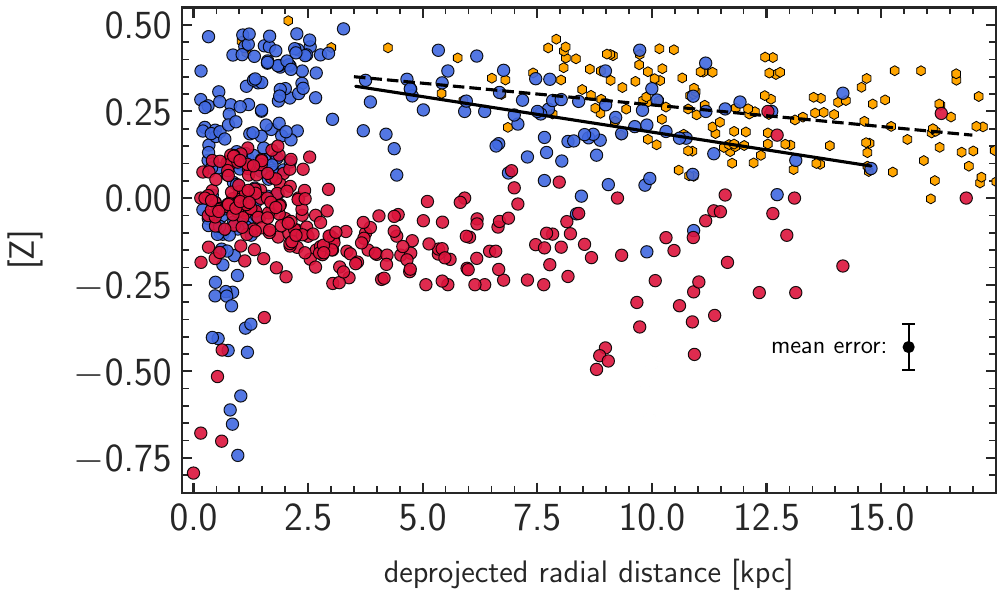} \medskip
	\caption{Radial distribution of the logarithm of metallicity relative to the sun. The young and old population are plotted in blue and red, respectively. HII-region metallicities based on oxygen abundances \citep{Ho2017} are shown in yellow. We also show regressions for the young stellar population (solid) and the HII-regions (dashed) calculated for galactocentric distances larger than 3.5 kpc.}     \label{fig:metgrad}
\end{figure*}

\newpage

\section{Analysis Method} \label{sec:analysis}

Our population synthesis method is described in detail in \citet{Sextl2023}. We fit the SED and absorption line spectra of the integrated stellar population with a linear combination of the spectra of single stellar populations (SSP) of different ages and metallicities. This allows us to constrain the average metallicity and age of the population together with reddening E(B-V), extinction A$_V$ and the ratio R$_V$ = A$_V$/E(B-V), which characterizes the steepness of the reddening law. The observed and SSP template spectra are normalized in the range between 5500 to 5550 \AA. Therefore, the metallicities and ages obtained in this way are V-band luminosity weighted averages (see \citealt{Sextl2023}). 

The model spectrum of the integrated stellar population M$_{\lambda}$ combines the spectra of single stellar populations (SSPs) f$_{\lambda, i}$(t$_{i}$, [Z]$_{i}$)  with age t$_{i}$ and logarithmic metallicity [Z]$_{i}$ = log Z/Z$_{\sun}$

\begin{equation}
  M_{\lambda} = D_{\lambda}(R_{V}, E(B-V)) \left[ \sum_{i=1}^{n_{SSP}} b_{i} f_{\lambda, i} (t_{i}, [Z]_{i}) + b_{a}f_{\lambda}^{a} \right]
\end{equation}  

where the coefficients b$_i$ describe the contribution of burst i with age t$_{i}$ and metallicity  [Z]$_{i}$ and D$_{\lambda}(R_{V}, E(B-V))$ accounts for the absorption by interstellar dust. b$_a$ accounts for the contribution of a featureless AGN continuum f$_{\lambda}^{a}$ with wavelength slope $\lambda^{-0.5}$ (as in \cite{Cardoso2017}). This additional template is only utilized in the fitting process of bins near the center which show broad-line region (BLR) features in their spectra.

We apply the Flexible Stellar Population Synthesis code (FSPS, version 3.2) \citep{Conroy2009, Conroy2010} for the calculation of the individual SSP spectra together with the MILES library \citep{Sanchez2006} and MESA stellar evolution isochrones \citep{Choi2016, Dotter2016} and a \citet{Chabrier2003} initial mass function are adopted. It is important to note that FSPS utilizes not only one main stellar library option, but adds supplementary stellar spectra from complementary evolutionary phases. Thus, the limited number of stars with T $ > 9000$ K in the empirical MILES library (as can be seen in \citet{Martins2007}), is enlarged by hot star spectra from \citet{Eldridge2017} and Wolf-Rayet spectra from \cite{Smith2002}. Spectra of AGB-, post-AGB- and carbon stars are added \citep{Lancon2000, Rauch2003, Aringer2009}. The final set of our SSP spectra is then adjusted to the TYPHOON spectral resolution. Because of the low TYPHOON spectral resolution we do not consider the line broadening effects of stellar velocity dispersion, which has been measured, for instance, by \citet{bittner2020} or \citet{Pessa2023}.

As in \citet{Sextl2023} we account for finite time lengths of the stellar bursts of 0.1, 1.0 and 10.0 Myr but we find that SSP spectra with the shortest burst length result in the best spectral fits with lowest residual $\chi^2$ value. To account for the effects of interstellar dust, we apply the attenuation law by \citet{Calzetti2000} with variable R$_V$ (deviating from the Calzetti standard value R$_V$=4.05), derived empirically in local starburst galaxies. 

For the choice of SSPs we use the high resolution age described in \citet{Sextl2023}, which starts at 0.1 Myr with a step to 1.0 Myr and then continues with logarithmic steps $\Delta$log t (in Gyr) alternating between 0.05 and 0.1 dex until 12.59 Gyr. The metallicities start at [Z] = -1.5 and increases in steps of 0.25 dex until [Z] = 0.5 as the highest value of the grid. We have, thus, a grid with n$_a$ = 52 ages and n$_z$ = 9 metallicities and a total number of n$_{SSP}$ = 468 SSPs.

In our fitting procedure we correct for the radial velocity shifts of the observed spectra. We also measure the equivalent widths of the ISM nebular emission lines of hydrogen H$_{\beta}$ to account for nebular emission continuum if needed (see \citealt{Sextl2023}). The equivalent width is used to estimate the contribution of the nebular continuum at the wavelength of H$_{\beta}$ and the wavelength dependence is calculated by accounting for nebular hydrogen and helium bound-free, free-free and 2-photon emission. The (generally weak) nebular continuum contribution is then subtracted from the observed spectra. In addition, as in \citet{Sextl2023} spectral regions contaminated by ISM emission or absorption lines are not included in the spectral fit of the stellar population.

After these steps, the coefficients b$_i$ and b$_a$ are then determined by adopting a grid of R$_V$ and E(B-V) values. For each pair of these quantities we calculate D$_{\lambda}(R_{V}, E(B-V))$, use a least square algorithm to directly solve for the coefficients b, calculate the model spectrum and a $\chi^2$ value by comparing with the observed spectrum. The minimum of $\chi^2$ defines the best fit. Errors are estimated by fitting the observed spectra modified by adding Monte Carlo Gaussian noise with zero mean and a standard deviation corresponding to the flux  error at each wavelength point. The uncertainties of the stellar population parameters are then calculated as the standard deviation of their distributions produced by 20 such Monte Carlo realizations.

Following the arguments in \citet{Sextl2023} we use a characteristic age limit of 1.6 Gyr to distinguish between the young and old population and determine average ages and metallicities of these two populations separately. Through t$_i \le$ 1.6 Gyr and t$_i \ge$ 1.6 Gyr, respectively, we introduce the young and old population and define the corresponding metallicities [Z]$_{y}$, [Z]$_{o}$ and ages log(t$_{y}$), log(t$_{o}$) via

\begin{equation}
  b_{y} = \sum_{i_{y}} b_{i} ,  b_{o} = \sum_{i_{o}} b_{i}
\end{equation}  

and

\begin{equation}
  [Z]_{y} = { \frac{1}{b_{y}} } \sum_{i_{y}} b_{i} [Z]_{i}
  \end{equation}
  \begin{equation}
  [Z]_{o} = { \frac{1} {b_{o}}} \sum_{i_{o}} b_{i} [Z]_{i}
  \end{equation}
  \begin{equation}
  log(t_{y}) = {\frac{1} {b_{y}}} \sum_{i_{young}} b_{i} log(t_{i})
  \end{equation}
  \begin{equation}
  log(t_{o}) = {\frac{1} {b_{o}}} \sum_{i_{old}} b_{i} log(t_{i}).
\end{equation}  

The average values of the total population, young and old, are then calculated as

\begin{equation}
  [Z] = b_{y}[Z]_{y} + b_{o}[Z]_{o}  ,  log(t) = b_{y}log(t_{y}) + b_{o}log(t_{o})
\end{equation}  

As explained in \citet{Sextl2023} metallicities and ages obtained in this way are luminosity weighted averages. The results of our analysis are presented in the next section.

\begin{figure}
        \smallskip
	\center \includegraphics[width=1\columnwidth]{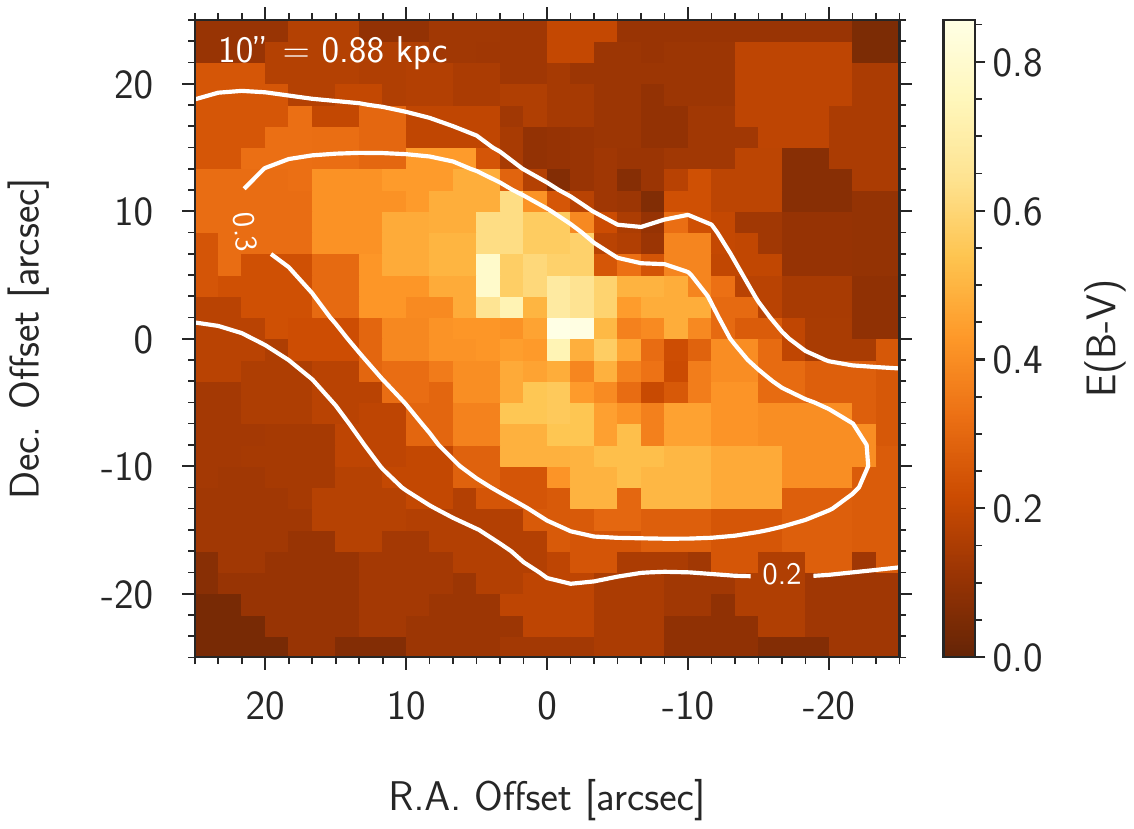}\medskip
    \center \includegraphics[width=1\columnwidth]{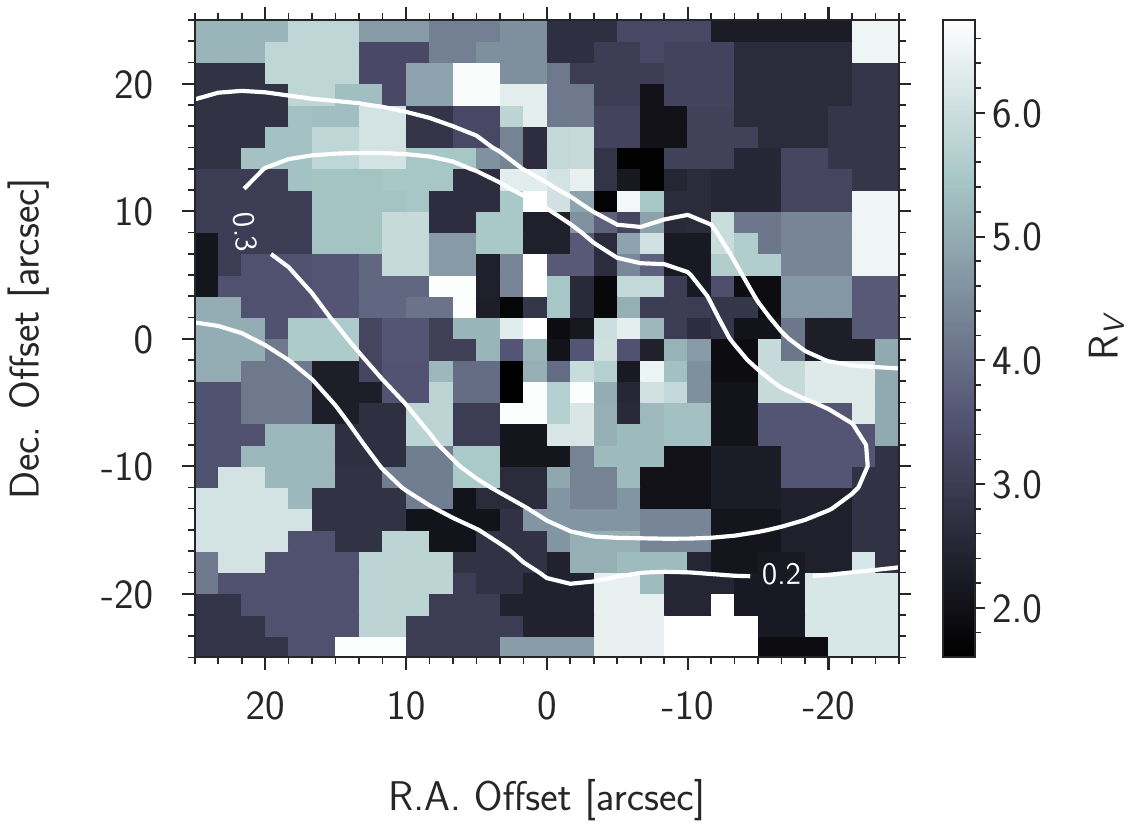}\medskip
    \center \includegraphics[width=1\columnwidth]{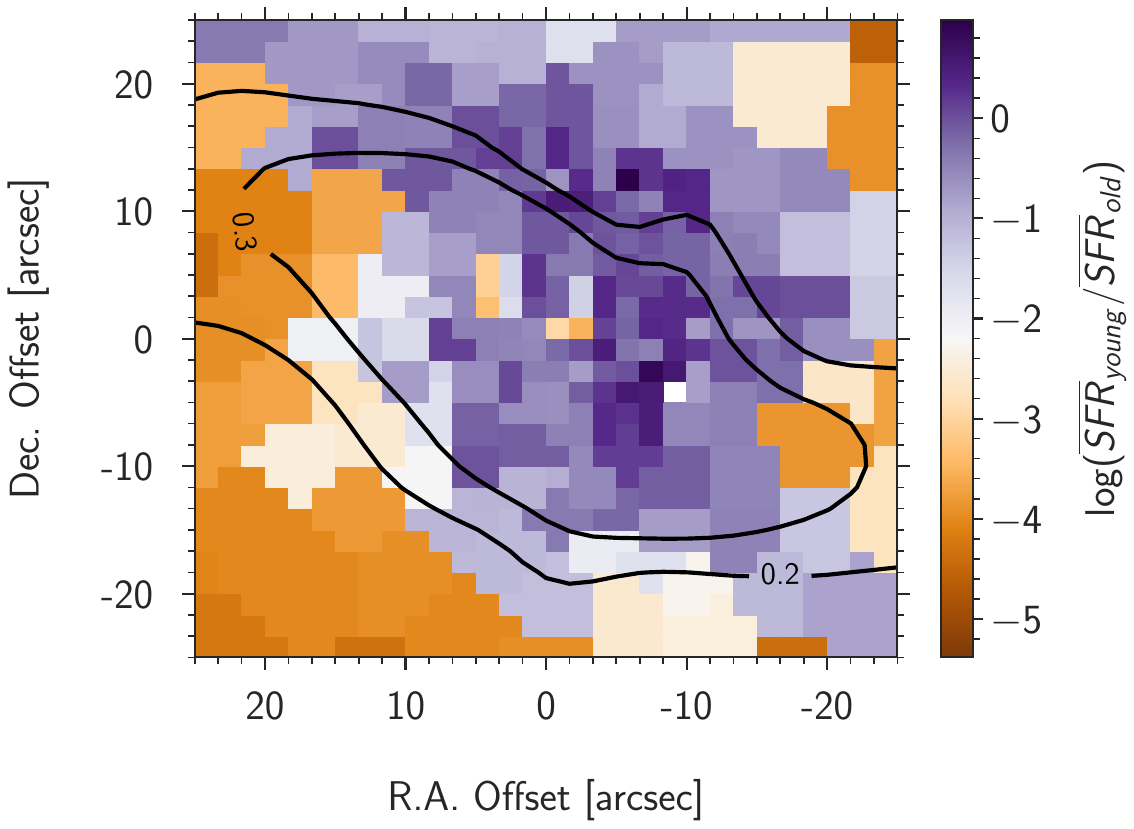}\medskip
	\caption{The central region of NGC 1365. Maps of the color excess E(B-V) (top), $R_{V}$ (middle), and the ratio of the average star formation rate of the young and old population (bottom). Isocontours of interstellar reddening E(B-V) are overplotted for E(B-V) = 0.2 and 0.3 mag.}     \label{fig:centralav}
\end{figure}


\begin{figure}
	\center \includegraphics[width=1\columnwidth]{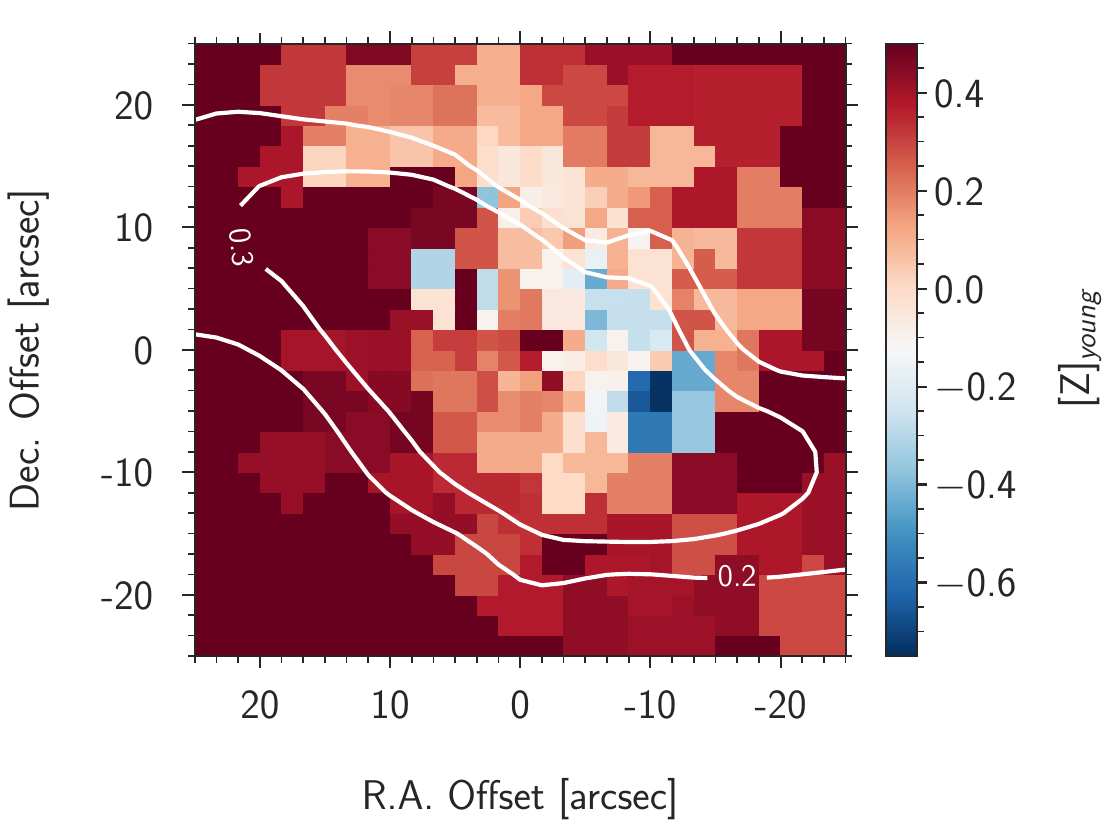}\medskip 
    \center \includegraphics[width=1\columnwidth]{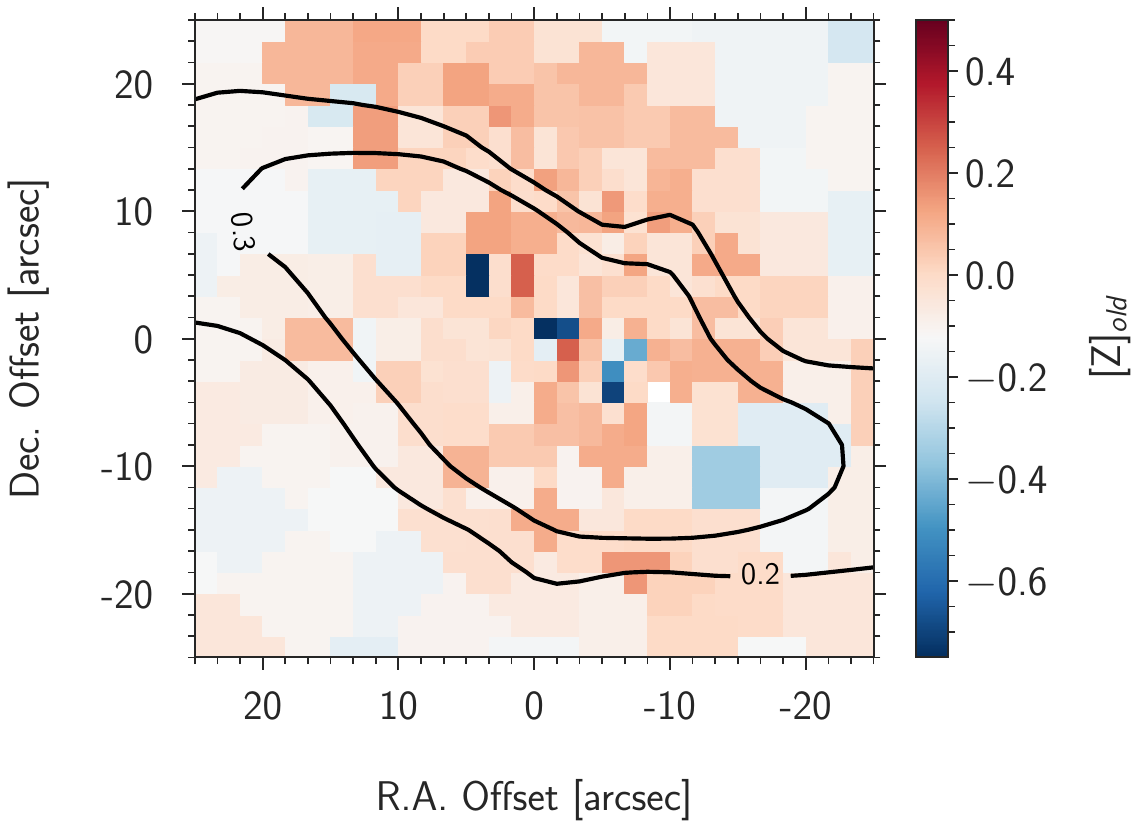}\medskip
    \center \includegraphics[width=1\columnwidth]{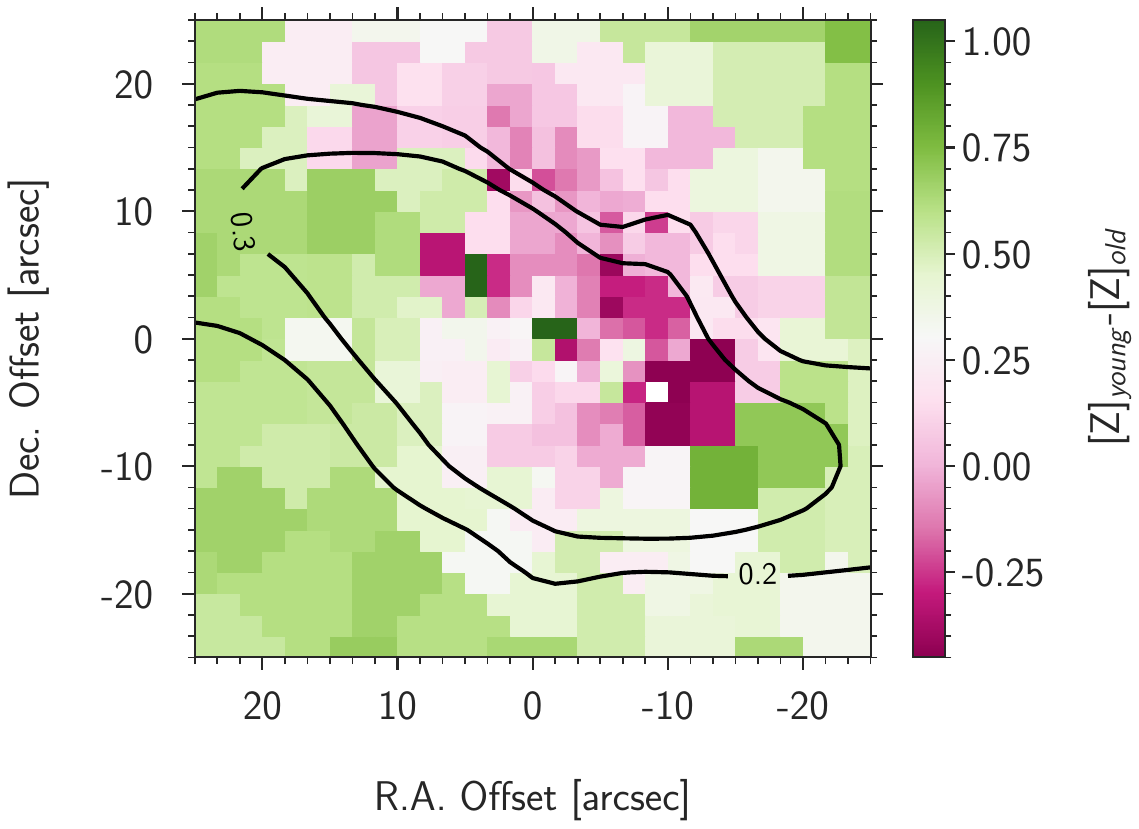}\medskip
	\caption{Central maps of stellar metallicity [Z]$_y$ and [Z]$_o$ of the young (top) and old (middle) population with the same axis range. The bottom figure shows the difference $\Delta$ [Z] = [Z]$_y$ - [Z]$_o$ . Negative differences are marked in pink tones. Isocontours of interstellar reddening E(B-V) are overplotted as in Figure \ref{fig:centralav}. The mean errors of [Z]$_y$ and [Z]$_o$ in the FoV are $\sim$0.12 dex and $\sim$0.10 dex, respectively.}     \label{fig:centralZ}
\end{figure}

\begin{figure}
        \medskip
	\center   \includegraphics[width=1\columnwidth]{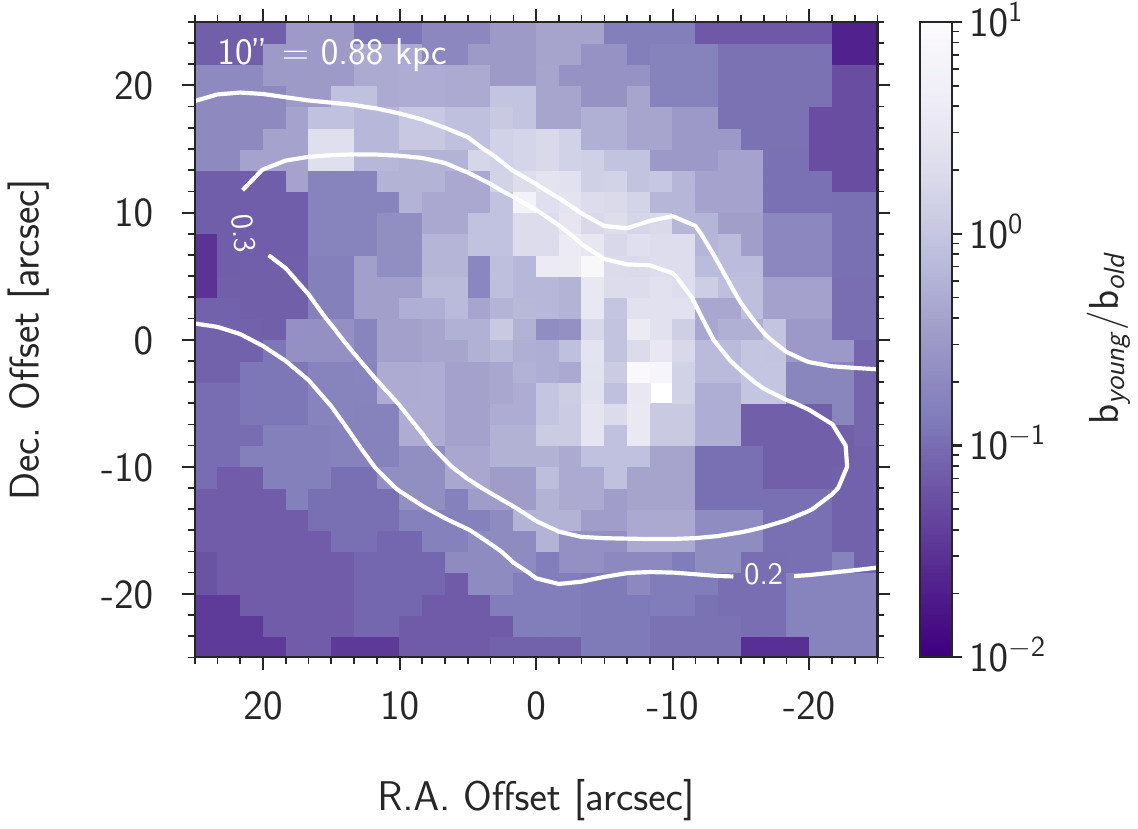}\medskip
      \center    \includegraphics[width=1\columnwidth]{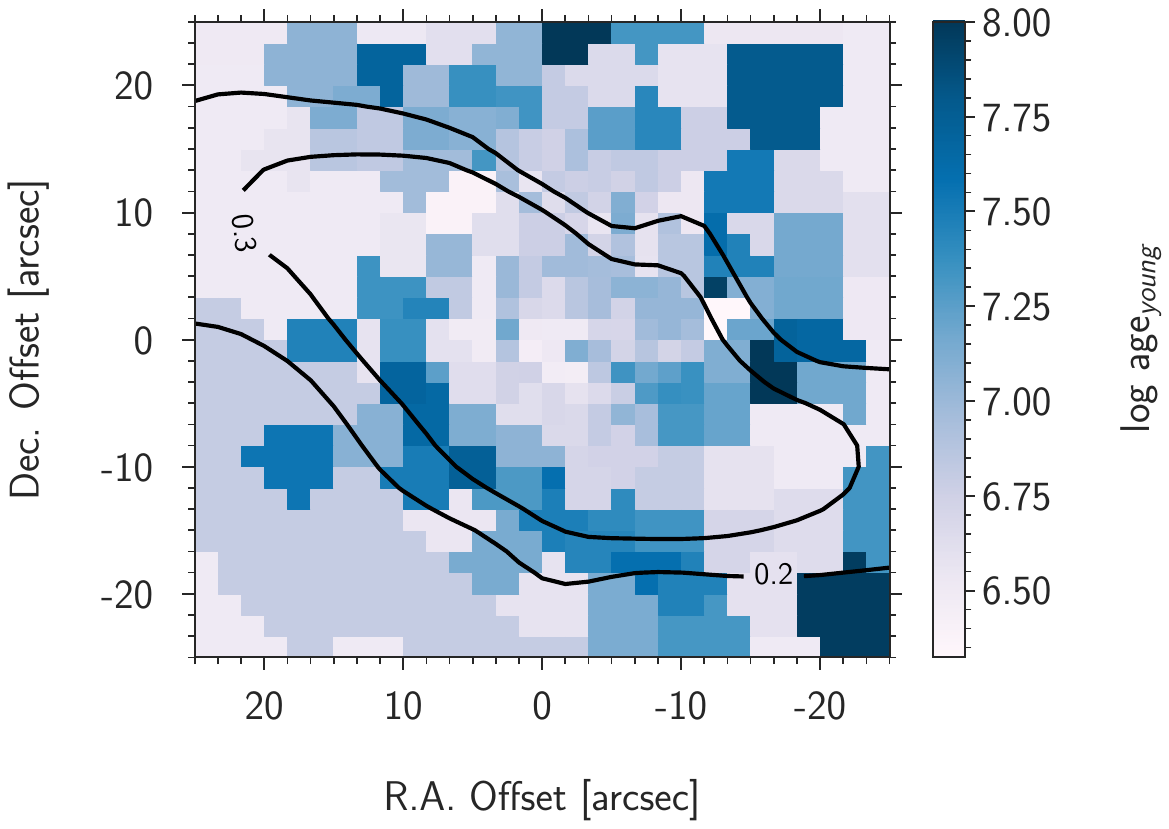}\medskip
       \center   \includegraphics[width=1\columnwidth]{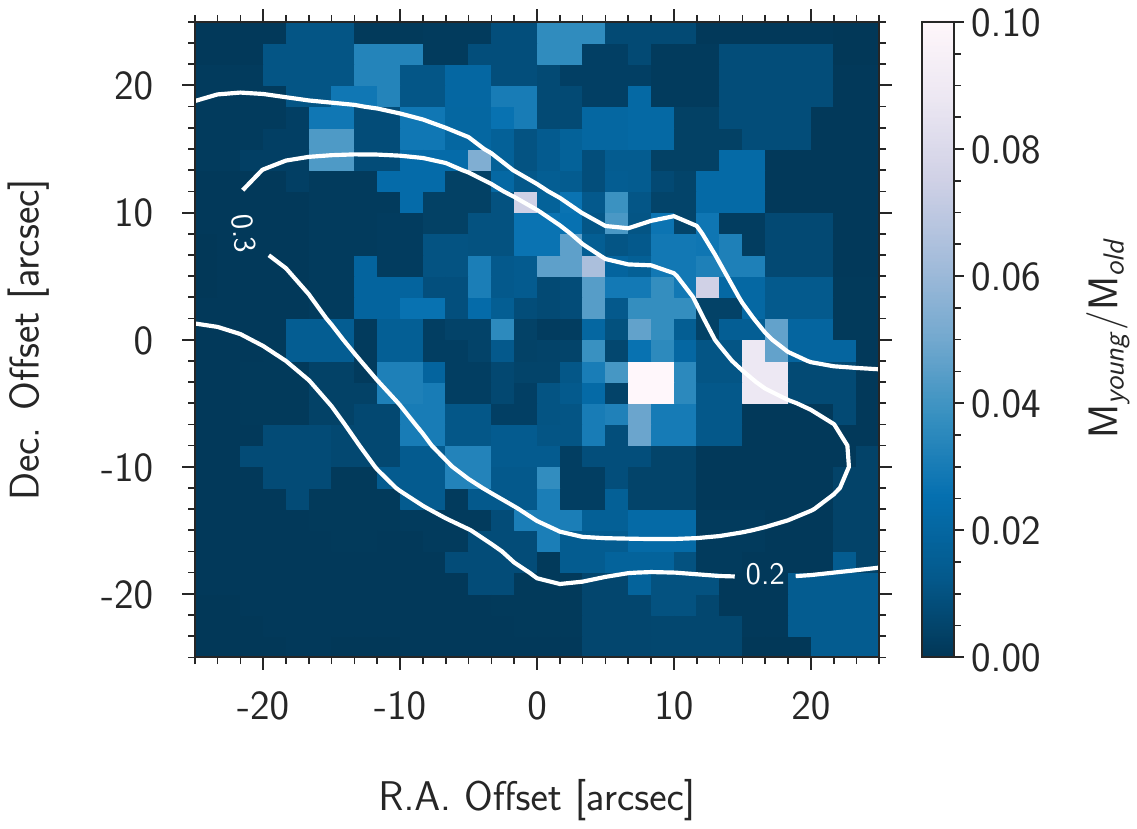}\medskip
	\caption{Central maps of the ratio b$_y$/b$_o$ (top), the ages t$_y$ of the young stellar population (middle), and the stellar mass ratio of the young and old population (bottom). Isocontours of interstellar reddening E(B-V) are overplotted as in Figure \ref{fig:centralav}.  }    
 \label{fig:centralage}
\end{figure}

\section{Results} \label{sec:results}

In the following subsections we present the results of our stellar population fits. We start with the maps of interstellar extinction. We then discuss the distribution of ages and metallicity of the young and old stellar population in the outer disk and the central region of the galaxy.

\subsection{Reddening and extinction} \label{subsec:extinction}

The spatial distribution of reddening E(B-V) and extinction A$_V$ by interstellar dust is shown in Figure \ref{fig:redden}. We find very strong effects in the center and bar and a moderate enhancement along the spiral arms. For the outer disk outside the spiral arms reddening is low and close to the Milky Way foreground reddening of E(B-V)$_{MW}$ = 0.018 mag \citep{Schlafly2011}. We provide additional information in Figure \ref{fig:redgrad}, which shows the radial distributions as a function of galactocentric distance together with the fit uncertainties. We note again that in our spectral fit of interstellar dust we have included the determination of R$_V$ which is allowed to deviate from the "standard" values such as 3.1 for the diffuse Milky Way ISM or 4.05 for the Calzetti-law in starburst galaxies. Such deviations are common in star forming galaxies (see discussion and references in \citealt{Sextl2023}). The bottom of Figure \ref{fig:redgrad} shows the radial distribution of the R$_V$ values which we encounter together with their errors. Bins with E(B-V) values $<0.005$ have been removed since the slope of the attenuation curve cannot be quantified in this case. We find a mean value around 4.0 but with a wide scatter between 1.5 and 6.7. As comprehensively discussed in \citet{Salim2020}, extinction is not the same as attenuation due to the additional geometric and scattering components in the latter case. This also implies that one should be cautious here with the common interpretation of large values in $R_{V}$ implying larger grain sizes \citep{Battisti2017, Calzetti2000}. It can nevertheless be used as an indicator for obscuration \citep{Calzetti2001}.

A more detailed discussion of the central region will be given below.

\subsection{Population ages and inside-out growth} \label{subsec:ages}

Figure \ref{fig:agemap} displays the spatial map of the average population ages across the surface of NGC 1365 and Figure \ref{fig:agegrad} (top) provides the corresponding radial distribution. We find a maximum of populations ages around a galactocentric distance of 5 kpc and lower ages towards the center and the outer edge. The decrease of age beyond 5 kpc is a clear indication that the outer disk is more and more dominated by the young stellar population. This is confirmed by the radial dependence of the ratio b$_{y}$/b$_{o}$, which represents the ratio of the luminosity contribution of the young and old population to the total observed population spectrum and which is also given in Figure \ref{fig:agegrad}. The contribution of the young population gradually increases when going beyond 3 kpc towards larger galactocentric distances. 

As shown in \citet{Sextl2023} (equations 11 and 12 and text at the end of section 4) the SSP fit coefficients b$_i$ can also be used to estimate the ratio of star formation rates of the young and old population. This ratio is also shown at the bottom of Figure \ref{fig:agegrad}. The outer disk shows a gradual increase of this ratio from 3 kpc outwards. We note that the average age of the young population in the different spatial bins of this region is between 0.1 and 1.0 Gyr with a very few exceptions of a very young population only 5 Myr old. The average ages of the old population are between 10 and 12.6 Gyr. 

The decrease of average ages and the increased contribution to luminosity and star formation by the young population towards the outer radii is a clear indication of inside-out growth of the stellar disk of NGC 1365. This will be further discussed in Section \ref{sec:discussion}.

From 3 kpc inwards the trend with galactocentric distance reverses. Average ages decrease and the contribution of the young population becomes stronger again. We will discuss this in the subsections below. 

\begin{figure}
	\center \includegraphics[width=1\columnwidth]{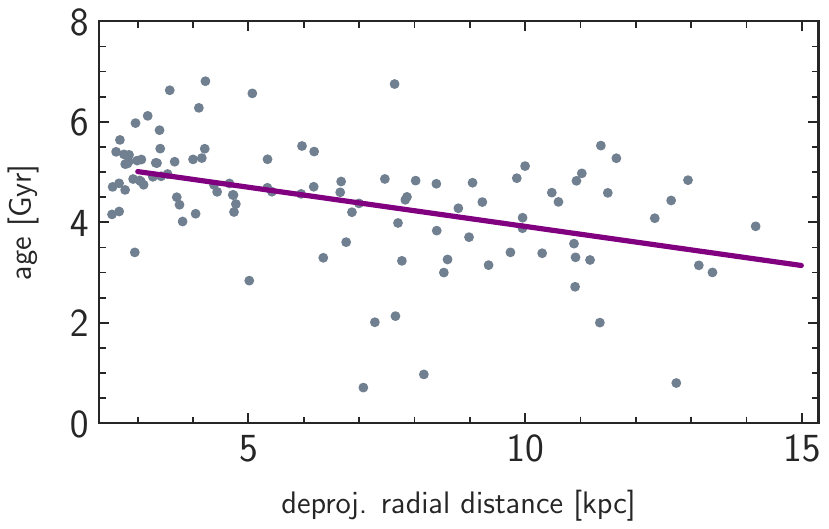}\medskip
	\caption{Age of the stellar population in each spectral bin at the time when 80 \% of the stellar population has formed as a function of galactocentric radius in kpc. A linear regression curve is also shown. For details see text. }   \label{fig:massgrowth}
\end{figure}

\subsection{The metallicity of the young and old population of the outer disk} \label{subsec:outer}

The radial distribution of metallicities (defined as [Z] = log Z/Z$_{\odot}$) for the young and old stellar population is plotted in Figure \ref{fig:metgrad}. We also add the metallicities of HII-regions obtained by \citet{Ho2017}. They are based on oxygen abundances as a proxy for metallicity and the Baysian strong line calibration developed by Ho et al. The \citealt{Asplund2009} value of log N(O)/N(H) + 12 = 8.69 has been adopted for the solar oxygen abundance.   

In the outer disk beyond 3.5 kpc we find a negative metallicity gradient of the young population of $-0.0207\pm0.0028$ dex/kpc in line with the expectations for a stellar disk which has formed inside-out. The stellar gradient is slightly steeper than the one obtained from HII-region emission lines and the value of metallicity is somewhat lower but these differences can be attributed to uncertainties of the HII-region strong line calibrations (see discussion in \citealt{Ho2017, Bresolin2016}). For the outer old population we note that the low metallicity bins with $[Z]\le -0.35$ at a distance of 8-11 kpc are all located within the leading arm of the southern spiral right outside the bar. Apart from this anomaly there is no clear indication of an outer radial gradient of the old population. This is a striking result and will be discussed in section \ref{sec:discussion}. The global difference of $\approx$~0.4 dex in [Z] between the young and old population agrees very well with the difference found for the massive star forming SDSS galaxies studied by \citet{Sextl2023} and with the prediction of chemical evolution models (see, for instance, \citealt{Kudritzki2021a,Kudritzki2021b}).

In addition to the outer negative gradient of [Z]$_y$ Figure \ref{fig:metgrad} reveals a steep drop to low stellar metallicities of the young population towards the center of the galaxy. The metallicity of the old population, on the other hand, increases on average. This will addressed in the next subsection where we discuss the central region of the galaxy.

\subsection{Central region} \label{subsec:central}

As already indicated in Figures \ref{fig:redgrad} and \ref{fig:agegrad} the central region of NGC 1365 is characterized by high interstellar extinction and a strong contribution of the young stellar population to the observed luminosity. This is demonstrated in more detail by Figures \ref{fig:centralav} and \ref{fig:centralage} which provide zoomed maps of the central region of reddening E(B-V), R$_V$, average star formation rate ratio, the luminosity contributions ratio of the young and old population, mass ratios, and the age of the young population. However, there is an obvious asymmetry in the sense that the increased luminosity contribution and star formation activity of the young population is confined mostly to the northern area with E(B-V) $\sim$ 0.2 to 0.3 mag and less is found in the corresponding south-east region. This has been discussed most recently by \citet{Schinnerer2023} in their analysis of JWST mid-IR imaging and ALMA CO observations. They concluded that the asymmetric gas infall along the bar has already initiated the intense star formation in the northern bar lane whereas in the dense molecular clouds in the south star formation has yet to come but will start very soon. This is in accordance with the rather similar gas properties in both regions.

We note that while luminosity contribution of the young population is very high in the central region, the ratio of masses of the populations is M$_{y}$/M$_{o}$ is small with a mean value of 0.03 (see Figure \ref{fig:centralage} bottom). The reason for this significant difference to b$_{y}$/b$_{o}$ is the higher luminosity of younger (and more massive) stars. Details of the computation of the mass ratio are described in Section \ref{sec:discussion}.

The central map of R$_V$ (Figure \ref{fig:centralav} middle) does not indicate any kind of spatial correlation. It appears that the R$_V$ values are randomly distributed depending on how different gas clouds contribute to the attenuation (see \citealt{Salim2020, Battisti2017}).

Most strikingly, as shown in Figure \ref{fig:centralZ}, the metallicity [Z]$_y$ of the young population formed in this central region is very low, sometimes even lower than the metallicity of the old population. Obviously, the surface bins with low metallicity in Figure \ref{fig:metgrad} are all confined to a coherent area in the very center. This population has been formed most recently, as can be seen from Figure \ref{fig:centralage}. This is in accordance with \citet{Whitmore2023} who identified several young star clusters with ages $< 10$ Myr in the central regions of NGC 1365. Since our spectral analysis contemplates all stars in a spaxel, not only the massive clusters, we expect slightly higher ages in our analysis. In general, it seems to be of prime importance to provide SSP templates with sufficient young ages for the fit to capture bins with such young clusters accordingly.

We note that central stellar metallicities of NGC 1365 have already been studied within two surveys using the ESO VLT integral field unit MUSE, the TIMER \citep{Gadotti2019, Gadotti2020, bittner2020} and the PHANGS \citep{Emsellem2022} survey. Both surveys do not distinguish between metallicities of the young and old population and provide only an average over the total population but also indicate a drop of total metallicity towards the center \citep{Pessa2023, Bittner2021}. From our result we learn that this drop is caused by a strong burst of the formation of a very young population of low metallicity.  

It is important to check whether the low metallicities encountered are an artefact of the fitting procedure or are caused by numerical uncertainty. As a first step, we have compared the  minimum fit $\chi^2$ values as a function of the metallicity obtained. We did not find any hint of a systematic difference. Fits at low metallicity have the same quality as high metallicity ones. In addition, we have applied an independent population synthesis algorithm and repeated the analysis for all TYPHOON bins by using the STARLIGHT package \citep{Cid_Fernandes2005, Asari2007} for all TYPHOON bins. Very similar results were found including the low metallicity bins in the center. We also note that keeping $R_{V}$ fixed to the Calzetti standard value of 4.05 or assuming a Milky Way reddening law with constant $R_{V}=3.1$ \citep{Cardelli1989} produces a similar peculiar region of low metallicity. 

We also experimented with differential interstellar extinction between the young and old population. Following \citet{LoFaro2017} and \citet{Yuan2018} we introduced a factor f$_{att}$, which increases the reddening of the very young stars (t $\le$ 10 Myr) relative to the older stars by adopting E(B-V)(t$\le$ 10Myr) = E(B-V)/f$_{att}$. Applying a value of f$_{att}$ = 0.5 we did not encounter substantial changes in the derived stellar properties larger than the fit uncertainties in the outer parts of the galaxy. However, in the central region, where the extinction is highest, the effect of inversion of metallicity between the young and old stellar population is increased. Fits with differential extinction tend to lower the metallicity for the young population even further by on average $\sim$ 0.15 dex whereas [Z]$_o$ remains the same. Since f$_{att}$ as an additional free parameter is poorly constrained, we refrain from a detailed follow up at this point but note that differential extinction tends to make the situation in the center more extreme.

Finally, we have checked how well the metal line absorption features are reproduced by our low metallity SSP fits. In Figure \ref{fig:lowzfit} we repeat the fit of bin 153 (already displayed in Figure \ref{fig:specfitT}) but now show the detailed fit of the metal absorption line features in two selected spectral regions. According to our fit the metallicity of the young population in this bin is [Z]$_y$ = -0.36. We conclude that given the quality of the spectra the metal lines are on average reproduced well. We take all this as a confirmation of our results. 

An obvious question is whether there are HII regions in this confined region of low stellar metallicity and what their metallicities are. Very unfortunately, the most recent PHANGS-MUSE \citep{groves2023} and TYPHOON (\citealt{Chen2023}) HII-region and ISM emission line studies do not provide sufficient information allowing for conclusions about the ISM metallicity in this specific region. This is mostly caused by the strong contribution of the central AGN to the ISM ionizing radiation field. 

One should also keep in mind that in the TYPHOON observation of NGC 1365 one spaxel corresponds to 145 pc, so the whole low-metallicity region covers approx. a circle of 2 kpc in diameter. In kpc-resolution surveys at higher redshifts such relatively small regions would have been overlooked easily or would be dismissed as faulty spaxels. This makes the local TYPHOON sample especially valuable.

The physical nature of the low metallicity of the central young stellar population is discussed in the next section.

\begin{figure}
        \medskip
	\center  \includegraphics[width=1\columnwidth]{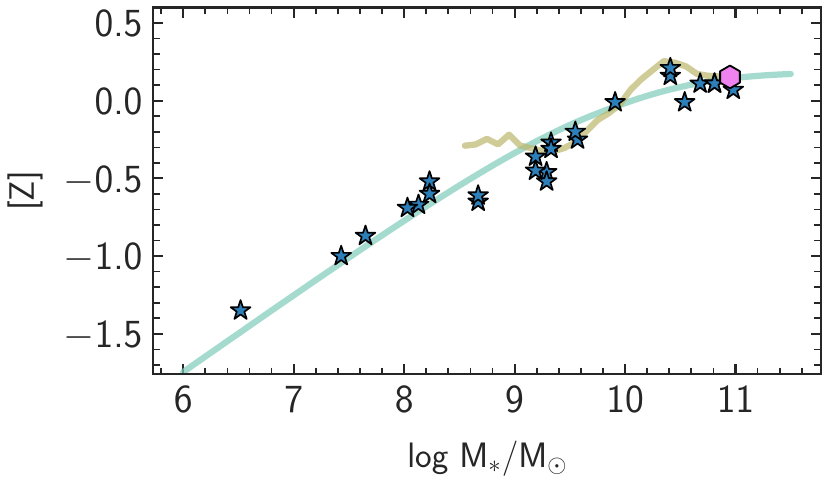}\medskip
	\caption{The mass metallicity relationship of star forming galaxies. Results from spectroscopy of individual red and blue supergiant stars are shown as stars (\citealt{Bresolin2022}, Urbaneja et al., 2023). These [Z] values refer to galactocentric distances of 0.4 R$_{25}$ for galaxies with a metallicity gradient and to mean values for low mass irregular galaxies without a gradient. The value for NGC 1365 is given by the pink hexagon (see text). The population synthesis results by \citet{Sextl2023} for the average metallicity of the young stellar population of SDSS galaxies are plotted as khaki lines}. The green curve is a prediction from galaxy evolution lookback models by \citet{Kudritzki2021a,Kudritzki2021b}.   \label{fig:MZR}
\end{figure}
\begin{figure}
        \medskip
	\center  \includegraphics[width=1\columnwidth]{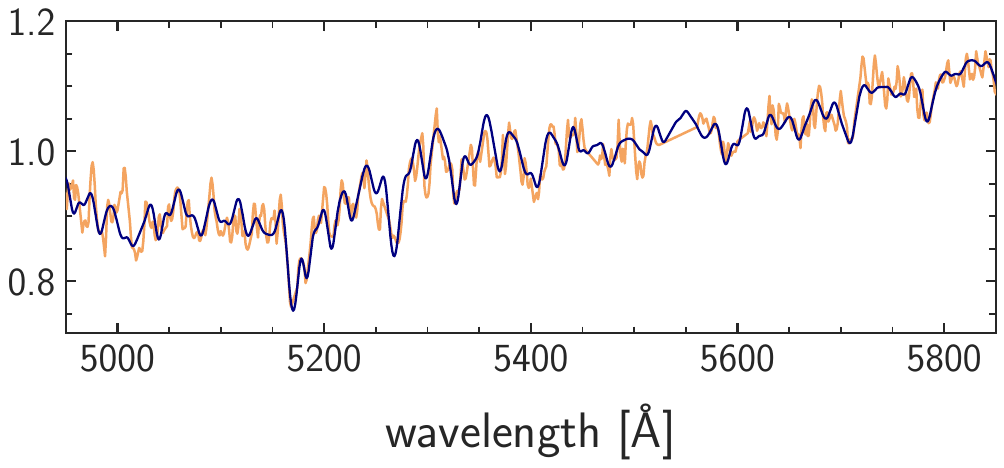}\medskip
    \center  \includegraphics[width=1\columnwidth]{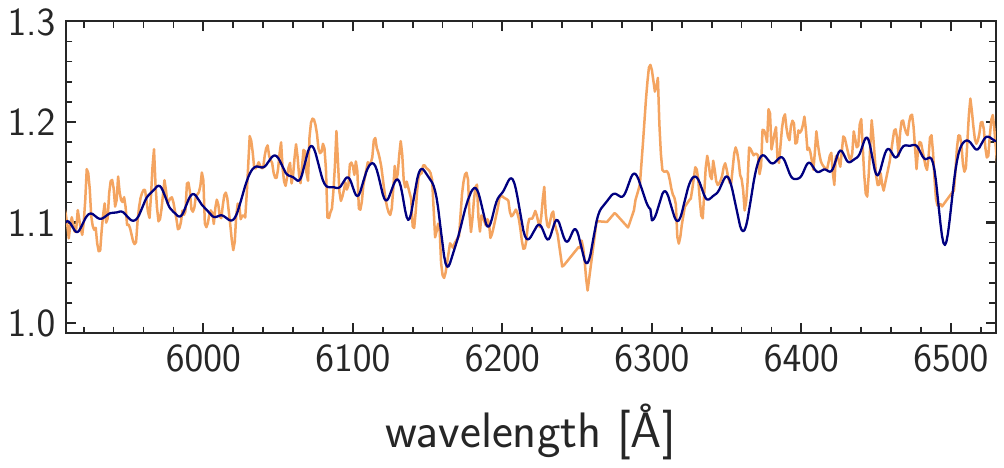}\medskip
	\caption{Spectral fit (black) of the absorption line features of bin 153. The metallicity of the young population is [Z]$_y$ = -0.36. Note that the TYPHOON spectrum has a gap from 5524 to 5561\AA. The spectral regions at 5007 and 6300 \AA~are not included in the $\chi^{2}$ spectral fit due to the contamination by ISM emission.}   \label{fig:lowzfit}
\end{figure}

\section{Discussion} \label{sec:discussion}

As concluded in section \ref{subsec:ages} and Figures \ref{fig:agemap} and \ref{fig:agegrad}, the decrease of ages averaged over the total population and the increase of the ratio b$_y$/b$_o$ is indicative of inside-out growth of the outer stellar disk of NGC 1365. An alternative way to investigate this is to look at the radial dependence of stellar mass growth. As described in \citet{Sextl2023} the fit coefficients b$_i$ in Equation (1) of the spectral fit of an observed spectral bin can be related to the relative number contribution N$_i$ of stars of isochrone i, for which the SSP model spectrum is calculated, and the V-band luminosity L$_i$(V) of the isochrone via

\begin{equation}
    b_i = N_{i}L_{i}.
\end{equation}

With N$_i$ = b$_i$/L$_i$ and the mass M$_i$ of each isochrone the function

\begin{equation}
    g(\log(t_i)) = \dfrac{\sum_{i_1=1}^{i} \sum_{i_2=1}^{n_z} N_{i_1, i_2} M_{i_1, i_2} } { \sum_{i_1=1}^{n_{a}} \sum_{i_2=1}^{n_z} N_{i_1, i_2} M_{i_1, i_2}}
\end{equation}

describes the accumulative mass growth in the surface area of each spectral bin as a function of time. Note that in Equation (9) the inner sums in the nominator and denominator add up the contributions by different metallicities at the same age, whereas the outer sums accumulate the contribution of different ages.

Applying Equation (9) we can calculate the age (or lookback time) for each spectral bin, when 80 \% of the stellar mass has formed. The corresponding radial distribution of ages is shown in Figure \ref{fig:massgrowth}. We see a clear indication of inside-out growth beyond 2.5 kpc. At the outermost parts of the disk the mass growth was accomplished about 2 Gyr later than in the inner parts. We note that a similar result has been obtained by \citet{Pessa2023}.

According to standard chemical evolution models including galactic winds and gas infall (see, for instance, \citealt{Hou2000, Chiappini2001, Kudritzki2015, Weinberg2017, Kang2021}) the inside-out growth of the disk is the physical reason for the negative metallicity gradient of the young stellar population and the ISM in star forming disk spiral galaxies, which we also encounter in NGC 1365. The value of -0.02 dex/kpc determined in our work seems very small but NGC 1365 is a huge galaxy and renormalizing with respect to the isophotal radius leads to -0.59 dex/R$_{25}$. This value is in good agreement with gradients found from the quantitative spectroscopy of individual supergiant stars in nearby galaxies (see, for example,  \citealt{Kudritzki2012, Bresolin2022, Liu2022} and references therein) and comprehensive studies of metallicity gradients obtained from HII-regions such as \citet{Ho2015}. 

The average outer metallicity of the young population represented by the value [Z]$_y$ = 0.15 at 0.4 R$_{25}$ = 11.82 kpc is in good agreement with the mass-metallicity relationship (MZR) of star forming galaxies. Figure \ref{fig:MZR} shows a MZR comparison with the results obtained by the analysis of individual supergiant stars (\citealt{Bresolin2022}, Urbaneja et al., submitted to ApJ) and the SDSS population synthesis study of the young population by \citet{Sextl2023}.

As already mentioned above, the average difference between the metallicity of the old and the young population of about 0.4 dex agrees with the result obtained by \citet{Sextl2023} in their investigation of the integrated spectra of 250000 SDSS galaxies. As for the metallicity gradients, the fact that we do not encounter a gradient for the outer old population is very interesting. When comparing gradients as a function of population age galaxy chemical evolution models yield different results depending on the assumptions made. \citet{Hou2000} find that gradients of the older population are steeper, whereas \citet{Chiappini2001} find the opposite. \citet{Roskar2008} point out that stellar migration will tend to flatten abundance gradients with time and predict younger populations to have a steeper gradient. It seems that population synthesis analysis might provide a powerful tool to investigate this issue and to provide additional constraints on galaxy evolution in this way. This will be an important part of our continued TYPHOON population synthesis survey.

A striking result is the low metallicity of the young population in a very confined region in the center (Figure \ref{fig:centralZ}). A straightforward first explanation is strong inflow of metal poor gas which has influenced the most recent star formation. This would be in line with the most recent work by \citet{Schinnerer2023} and the results by \citet{Sanchez2008} who detected gas infall along the bar. \citet{Tabatabaei2013} gauged a relatively short timescale of only 300 Myrs in their submilimeter analysis on which gas from the bar looses angular momentum and flows into the center. However, it is not clear whether this infalling gas is metal poor and whether the amount of gas accumulated has been sufficient to form a new metal poor population. In principle, galaxy evolution models with infall and outflow can produce a younger population with a metallicity lower than the older population \citep{Spitoni2015} but the differences are not as extreme as we encounter in NGC 1365. A crucial test would be a metallicity measurement of the infalling gas. 

We propose an additional scenario as an explanation for the low metallicity of the central young stellar population, namely interrupted chemical evolution. In this scenario chemical evolution in the central region started in the usual way with heavy star formation and building up metals. Then, at a time when ISM metallicity had already reached high values of above solar, feedback from the central AGN together with stellar feedback from supernovae and winds created a strong central galactic wind and ejected the central gas so that star formation stopped for a long time. For a massive galaxy such as NGC 1365 an AGN can stay intermittently active for 1-5 Gyr and inhibit star formation on a scale of 1 Gyr \citep{Stasinska2015}.

In the meantime, the most massive stars which have already formed, continue to recycle metal enriched gas to the ISM through supernovae and winds but due to the energetic AGN and stellar activity this matter continues to be expelled. Then, after several Gyr this activity drops and a new reservoir of metal poorer ISM gas is built up again from stellar winds and mass-loss of stars of low mass of the older population, which formed earlier. This central reservoir eventually becomes dense enough to start formation again and the newly formed stars show the low metallicity composition of the older population.

\begin{figure}
        \medskip
	\center \includegraphics[width=1\columnwidth]{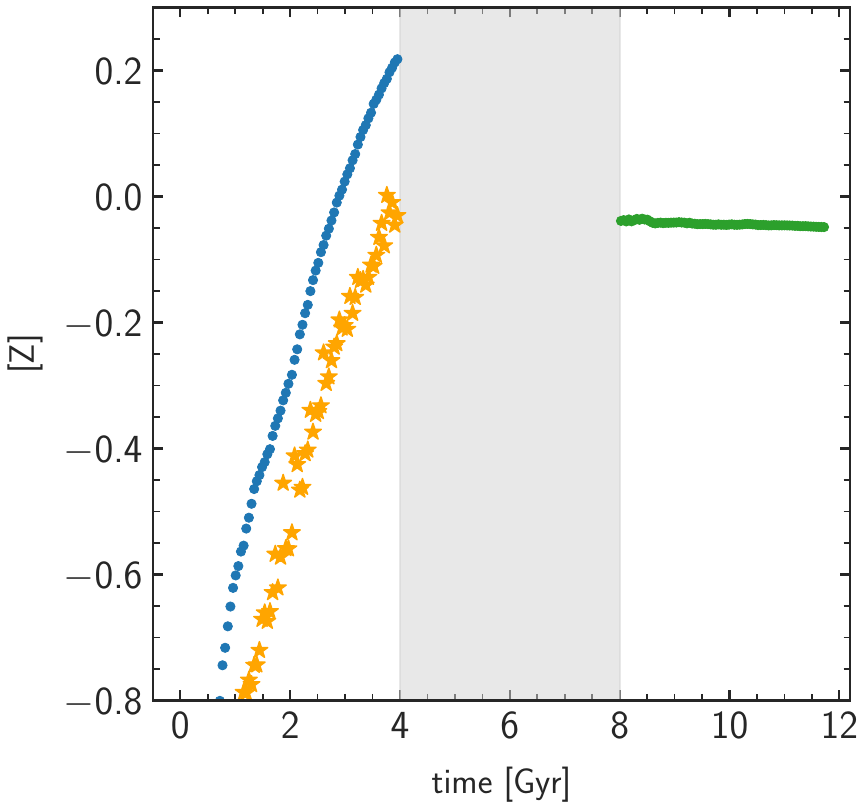} \medskip
	\caption{A simple chemical evolution model as example for the central region of NGC 1365. The metallicity of the ISM present in the center before the end of star formation at 4 Gyr is given in blue and the V-band luminosity averaged metallicity of the stars in orange. The abscissa is the evolution time in Gyr. The gray shaded area marks the time when the central ISM gas is ejected and star formation is quenched. The green symbols show the gas metallicity provided by stellar winds and supernovae of the old population after the central activity ejecting gas has stopped. For more details see text.}  \label{fig:chemevo}
\end{figure}

Figure \ref{fig:chemevo} shows the result of a simple Monte Carlo simulation, where we apply a closed-box chemical evolution model. We assume that stars form out of pristine gas with zero metallicity in an intense star formation process following a standard initial mass function \citep{Kroupa2001}. Metal enriched gas is recycled to the ISM through supernovae and stellar winds. We distinguish between stars with masses higher or lower than 9M$_{\odot}$. For the former, all matter except a remnant of 3M$_{\odot}$ is returned to the ISM and the mass of the convective core \citep{Maeder1987, Ekstroem2012} is converted into metals by core-collapse supernovae. For the latter, we adopt a white dwarf mass using an initial-final mass relationship \citep{Cummings2018} and recycle the remaining matter. For the metals produced by these objects we follow \citet{Kobayashi2009} and assume that 7 percent of the stars in the mass range between 3 to 9 M$_{\odot}$ increase the mass of the final WD through binary accretion until the Chandrasekhar limit. A subsequent Supernova type Ia explosion returns the white dwarf mass as metals. For the stellar life times of both groups we use the main sequence fit values given by \citet{Ekstroem2012}. In the case of SNIa explosions we do not account for the accretion time, which is short compared with the stellar life time \citep{Kobayashi2009}.

As a result of our calculation, ISM metallicity (shown in blue) is increasing quickly. Young, newly formed stars will resemble the ISM metallicity but the metallicity of older stars will be lower. We show the V-band luminosity averaged metallicities of the stars in orange. (As mentioned above the metallicities and ages we derive in our population synthesis approach are luminosity weighted quantities). 

After 4 Gyr star formation comes to a hold because of the energetic processes in the center, which eject all star forming gas, and the evolution stops. After all the ISM gas is quickly expelled from the center, the existing massive stars still recycle gas to the ISM. However, because of the AGN activity the new ISM gas still continues to be expelled and the central region remains gas free. Then, 8 Gyr after the whole chemical evolution started, the central activity stops and a new reservoir of ISM gas forms because of stellar winds and mass-loss from the existing low mass stars of different ages and low metallicity. The ISM metallicity of this new reservoir of gas is shown in green. It is slightly lower than the luminosity weighted metallicities of the old stars, which agrees qualitatively with Figure \ref{fig:centralZ}. Young stars forming out of this gas will resemble this metallicity. This explains how the formation of a lower metallicity young stellar population in the center is possible without invoking the infall of metal poor gas from outside.

An obvious question is whether the dying stars of the old population can provide enough gas for a new young population. As a very simple estimate we consider stars born at 3 Gyr in Figure \ref{fig:chemevo}. In our scenario the gas provided by stars with a main sequence lifetime shorter than 5 Gyr (or more massive than 1.18 $M_{\odot}$) is expelled by the central AGN activity. However, the gas recycled by stars less massive than 1.18 $M_{\odot}$ but with a higher mass than 0.958 $M_{\odot}$ (corresponding to a main sequence lifetime of 9 Gyr) can settle into the central region, because AGN activity has now stopped (we adopt $\tau_{MS} = 8(M/M_{\odot})^{-2.8}$ Gyr for the low mass main sequence lifetime). Using the IMF we calculate the mass of these dying stars in relation to the mass of all remaining stars with masses lower than 0.958 $M_{\odot}$ and obtain a ratio of 0.124. Assuming that half of this mass fraction is recycled as gas we obtain a ratio of gas mass to the mass of old stars of 0.06, which is a factor of two higher than the observed average mass ratio of the young to the old population (see Figure \ref{fig:centralage} bottom). A starburst of very high star formation efficiency \citep{Fisher2022} could, thus, explain the observations within the framework of our model.

The goal of this simplified calculation is only to demonstrate that such an additional scenario can work. A better match of the observations could certainly be obtained by modifying parameters such as star formation rates, evolution and yields etc. and, most importantly, by including infall of metal poor gas, but that is beyond the scope of this paper. We note that the idea of recycled gas accumulating from an aging stellar population is not new, see for instance in \cite{Ciotti2007}.

The situation in the center of NGC 1365 is not unique as it seems. The Milky Way shows a metallicity of the young stellar population which increases from [Z] $\sim$ -0.35 at 15 kpc galactocentric distance to $\sim$ +0.3 at 4 kpc \citep{Genovali2014,daSilva2022} but suddenly drops to solar or even subsolar towards the galactic center \citep{Najarro2004, Najarro2009, Davies2009a, Davies2009b, Martins2008, Origlia2013, Do2015}. Figure 5 in \citet{Genovali2014} gives an excellent impression of the very similar situation in the Milky Way. Additionally, recent ESO VLT observations revealed an intense star formation period in the galactic center $\sim$ 0.6-1 Gyr ago after a long quiescent phase of approximately 6 Gyrs \citep{Nogueras2020}.

In summary, we conclude that detailed spatially resolved population synthesis studies of young and old populations in star forming galaxies are an extremely powerful tool to investigate the evolution of galaxies. We will continue this work within the framework of our TYPHOON survey.

\begin{acknowledgments}
Acknowledgements. This work was initiated  and supported by the Munich Excellence Cluster Origins funded by the Deutsche Forschungsgemeinschaft (DFG, German Research Foundation) under Germany's Excellence Strategy EXC-2094 390783311.
\end{acknowledgments}

\bibliography{NGC1365_Apj}{}


\end{document}